\input{aipcheck}

\documentclass[
    ,final            
  ]
  {aipproc}

\layoutstyle{6x9}

\def \kms{\ifmmode{~{\rm km\,s}^{-1}}\else{~km~s$^{-1}$}\fi}
\def \vhel{\ifmmode{V_{{\rm hel}}}\else{$V_{{\rm hel}}$}\fi}
\def \vsys{\ifmmode{V_{{\rm sys}}}\else{$V_{{\rm sys}}$}\fi}
\def \degree{\ifmmode{^{\circ}}\else{$^{\circ}$}\fi}
\def \lsun{\ifmmode{{\rm\ L}_\odot}\else{${\rm\ L}_\odot $}\fi}
\def \msun{\ifmmode{{\rm\ M}_\odot}\else{${\rm\ M}_\odot$}\fi}
\def \myr{\ifmmode{{\rm\ M}_\odot{\rm\ yr}^{-1}}\else{${\rm\ M}_\odot$yr$^{-1}$}\fi}
\def \teff{\ifmmode{{\rm{T}}_{\rm eff}}\else{${\rm{T}}_{\rm eff}$}\fi}
\def \mdot{\ifmmode{{\rm\dot{M}}}\else{${\rm\dot{M}}$}\fi}

\newcommand{\ha}{H$\alpha${}}
\newcommand{\arcs}{$^{\prime\prime}${}}
\begin{document}

\title{Constraints on supernova progenitors from spatial correlations with \ha\ emission}

\classification{97.60Bw, 98.20.Af, 98.38.Hv}
\keywords      {galaxies:spiral-galaxies:photometry-galaxies:statistics-stars:supernovae:general}

\author{J.P. Anderson}{
  address={Astrophysics Research Institute, Liverpool John Moores University,
Twelve Quays House, Egerton Wharf, Birkenhead, CH41 1LD, UK\\
e-mail: jxa@astro.livjm.ac.uk}
}

\author{P.A. James}{
  address={Astrophysics Research Institute, Liverpool John Moores University,
Twelve Quays House, Egerton Wharf, Birkenhead, CH41 1LD, UK\\
e-mail: jxa@astro.livjm.ac.uk}
}

\author{M. Salaris}{
  address={Astrophysics Research Institute, Liverpool John Moores University,
Twelve Quays House, Egerton Wharf, Birkenhead, CH41 1LD, UK\\
e-mail: jxa@astro.livjm.ac.uk}
}

\author{S.M. Percival}{
  address={Astrophysics Research Institute, Liverpool John Moores University,
Twelve Quays House, Egerton Wharf, Birkenhead, CH41 1LD, UK\\
e-mail: jxa@astro.livjm.ac.uk}
}

\begin{abstract}
We have attempted to constrain the progenitors of all supernova types, through correlations of the positions
of historical supernovae with recent star formation, as traced by \ha\ emission. Through pixel statistics we
have found that a large fraction of the SNII population do not show any association with current star
formation, which we put down to a `runaway' fraction of these progenitors. The SNIb/c population accurately
traces the \ha\ emission, with some suggestion that the SNIc progenitors show a higher degree of correlation 
than the SNIb, suggesting higher mass progenitors for the former. Overall the SNIa population only show a weak
correlation to the positions of HII regions, but as many as a half may be associated with a young stellar population.
\end{abstract}

\maketitle


\section{1. Introduction}
Supernovae (SNe) are increasingly used in many areas of astrophysics, both as probes of physical processes in the local
Universe, and as cosmological probes out to high redshift. It is therefore extremely important that we tie down the progenitors of
these explosive events, in order to understand these objects individually while also increasing 
our understanding of a wide range of astrophysical problems.\\
The direct detection of SN progenitor stars in pre-explosion
images has been successfully achieved in a number of cases (see \cite{li05} and references therein), however 
this is only possible for core-collapse (CC) SNe in very nearby galaxies. Other approaches involve investigating how the
rates of the different types scale with host galaxy properties such as colour \cite{mann06}, or how the rates evolve with redshift
\cite{capp05}. Our approach is intermediate between these two methods, and involves a study of the stellar environment
local to the positions of historical SNe. Through this investigation we plan to constrain both the age and metallicity
of these parent stellar populations and hence further constrain the progenitors of all SN types.\\
Here
we summarise our initial progress on this research where we have used \ha\ imaging of galaxies, to investigate the association
of different types of SNe with a young stellar population.
\ha\ should accurately trace the most recent star formation (SF) within galaxies, a point that was made by Kennicutt \cite{kenn98}, and 
shows the utility of using \ha\ in the current context.
He commented that ``only stars with masses >10 \msun\ and lifetimes
of <20 Myr contribute significantly to the integrated ionising flux''. Thus,
\ha\ should plausibly trace the CC SN population within host galaxies.\\
Previous studies of this kind (\cite{vandyk96}, \cite{bart94})
have also investigated how the positions of CC SNe correlate with \ha\ emission and have 
found that most SNe lie near to HII regions, with a few exceptions lying as much as 40\arcs\ from their nearest region. The case for SNIa is less
clear. SNIa are thought to arise 
from a relatively older stellar population, so we would therefore not expect these SNe to show any association with recent SF. 
However, recent results \cite{mann06} suggest that there may be a fraction of the SNIa population that have relatively young progenitors. 
It is therefore interesting to investigate the possible association of this SN type with a young stellar population, with this in mind.\\
Thus our initial science questions were as follows:\\
(1) Do SNII and SNIb/c trace the same high mass stellar population? If not then what can this tell 
us about differences between the progenitors of these SNe? \\
(2) Do SNIa show any association with a young stellar population?\\
We build on the work of \cite{vandyk96} and \cite{bart94}, and produce a new statistical analysis, 
from which we present our results and draw our conclusions.

\section{2. Data}
The original database of observations for this study was the \ha\ Galaxy Survey (\ha GS), a survey of the local star formation properties of the Universe
undertaken with the Jacobus Kapteyn Telescope (JKT) 
(details can be found in \cite{james04}). The International Astronomical Union (IAU) database of SNe was searched to find all known 
SNe that had occurred throughout history in the 327 \ha GS galaxies. 63 such SNe were found, and the subsequent results and discussion 
from this sample can be found in \cite{mypaper06}. This sample has now been significantly increased by recent observations. $R$-band 
and \ha\ imaging have been obtained for a number of SN hosting galaxies with the Liverpool Telescope (LT)
and the Isaac Newton Telescope (INT), both situated on La Palma in the Canary Islands. The sample discussed here contains
51 SNII, 20 SNIb/c and 15 SNIa.\\
All pairs of $R$-band and \ha\ images were fully reduced and aligned at the sub-pixel level using standard reduction routines. The
$R$-band images were then scaled to the \ha\ images and used to continuum subtract the latter.
One of the most critical factors in our statistical analysis, is to obtain the correct SN containing pixel 
on the host galaxy images. Standard astrometric calibration was achieved, with typical fit residuals of 0\arcs .2  in each axis,  
using XDSS images downloaded from the 
Canadian Astronomy Data Centre website.\\
Our analysis has involved using a variety of statistical tests, including radial positions of SNe, galaxy luminosities and emission line fluxes, and pixel statistics.
A detailed analysis of these tests and their results (using the initial \ha GS sample) can be found in \cite{mypaper06}; here we concentrate on the
pixel statistics.\\   
Previous studies looking at the association of the positions of SN events in local galaxies (e.g. \cite{vandyk96}, \cite{bart94}), used  
the projected separation of each SN position from its nearest bright HII region. However, using this technique it can be hard to define the nearest HII region 
and therefore the distance to measure. We therefore present an unambiguous measure of the association of each SN with the \ha\ of its
host galaxy. \\
For each SN in our sample we have an associated \ha\ image of its host galaxy. The pixels on each image were sorted into increasing pixel count, starting from the most 
negative sky value, up to the most positive \ha\ flux pixel. Alongside this count we also formed the cumulative distribution of these pixel counts.  
A normalisation of this distribution was then achieved by dividing the sequence by the total
flux from the galaxy. All negative values were then set to zero, so that we have a normalised cumulative rank pixel 
function (NCR), that runs from 0 to 1, with one entry for every pixel on the array. Within this function, a value of 0 corresponds to zero \ha\ flux,
i.e. sky values, and a value of 1 corresponds to the centre of the brightest HII region of the host galaxy.
A detailed discussion of this statistic and its associated errors can be found in \cite{mypaper06}.\\
The SN-containing pixel obtained through our astrometry, was than used to give an NCR value for each SN in our sample.
The NCR function was produced so that if a SN progenitor type accurately traced the SF of its host galaxy, then one would expect a flat distribution of the 
NCR values, and a mean NCR value of 0.5.

\section{3. Results}
The mean NCR value for the 51 SNII in our sample is 0.212 (standard error on the mean of 0.035). Figure 1 shows the distribution of these values.
This histogram highlights the surprising result that the distribution is strongly weighted to low values, with around a third of the SNII population
lying on regions of zero SF. It is also interesting to note the lack of SNe coming from the centres of bright HII regions (high NCR values). This
may be a selection effect against detecting SNe in areas of high surface brightness, however this deficiency is not seen for either the SNIb/c or SNIa.\\
\begin{figure}
  \includegraphics[height=.24\textheight]{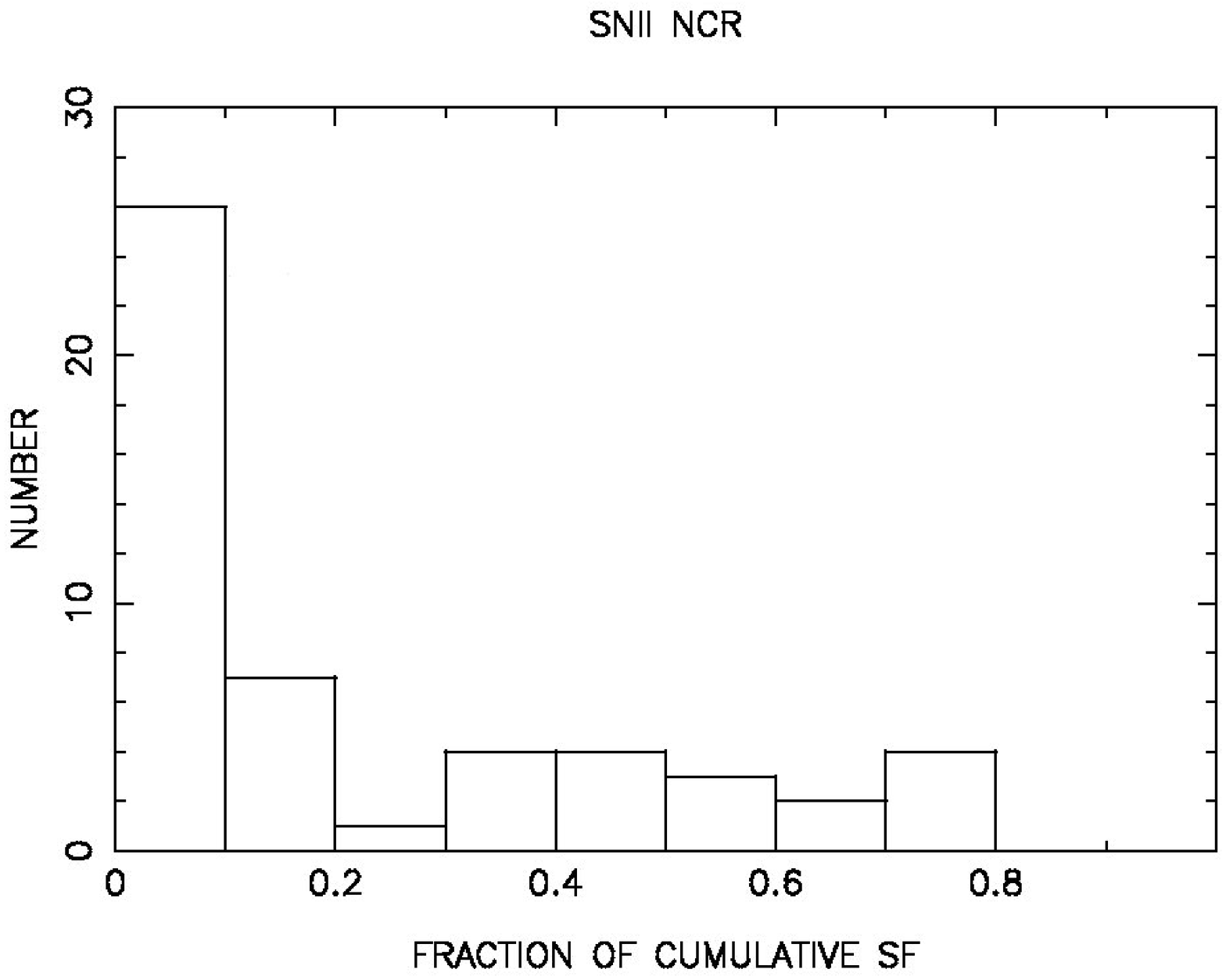}
  \caption{Histogram of NCR values of 51 SNII}
\end{figure}
We find a mean NCR value for the 20 SNIb/c of 0.536(0.097). This progenitor population seems 
to accurately trace the SF of the host galaxies, as would be expected for SNe with the highest mass progenitors. We can also split this population into 
the two subgroups. We then find a mean NCR value for the 7 SNIb of 0.379(0.097) and for the 11 SNIc of 0.588(0.113).
These results indicate that the SNIc progenitor population may be showing a stronger association to the 
\ha\ emission, implying higher mass progenitors.\\
The mean NCR value for the 15 SNIa is 0.247(0.083), and a histogram of the distribution of these values can be seen in Fig. 3. Here we find the intriguing
result that $\sim$1/2 of the progenitor population apparently show an association with the SF of their host galaxies, falling on pixels of non-zero \ha\ flux. 
Given the recent result of \cite{mann06}, predicting a bimodal progenitor population for SNIa, it is interesting to note the suggestion of this bimodality 
in our NCR values. We tested this using the DIPTEST statistic \cite{hart85}, and found tentative evidence 
that the distribution is bimodal with a probability of 50-90 $\%$.\\
\begin{figure}
  \includegraphics[height=.24\textheight]{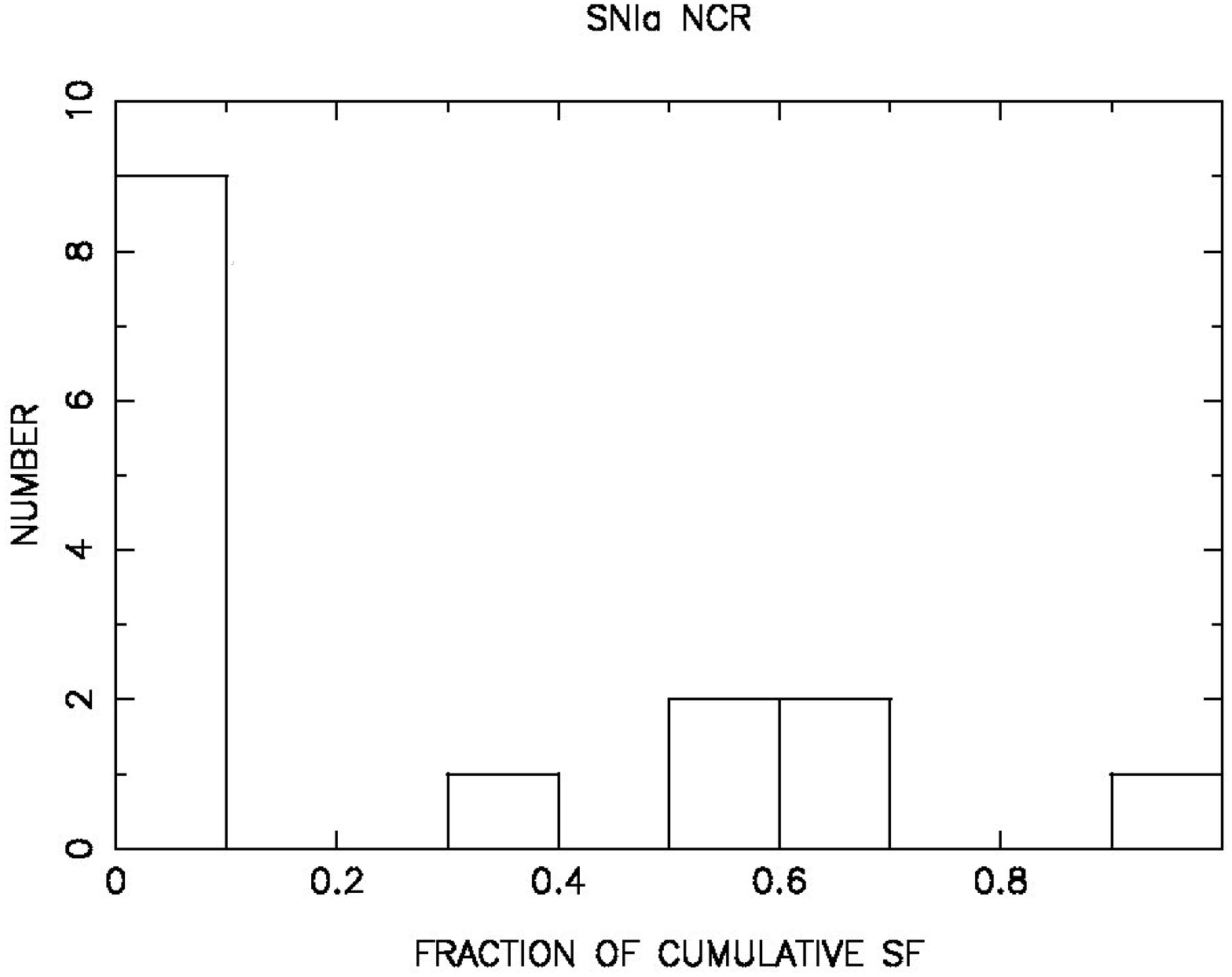}
  \caption{Histogram of NCR values of 15 SNIa}
\end{figure}

\section{4. Discussion and Conclusions}
Perhaps the most surprising result from this investigation is the non-association of a large fraction of the SNII progenitor population 
(assumed to have high mass progenitors), with the SF of their host galaxies. To date we put this
result down to a `runaway' fraction of the progenitors. Under this hypothesis these SNe 
did indeed originally reside within an HII region, but have since moved to their position of explosion between stellar
birth and SN. To test this hypothesis, the peculiar velocities that would be needed to move these SNe
from their nearest bright HII region to their position of explosion were calculated. 
The velocities calculated were in agreement for those of OB runaway stars. Thus this could be the explanation of this result.
However, this would seem to put the runaway fraction a quite a large value, and would also seem to rule out the binary-binary 
mechanism for runaways \cite{pov67}. Alternative explanations include that \ha\ is not an accurate tracer of
recent SF, or that these `runaway' SNe do actually explode within HII regions, but are obscured by dust, through which the SN light penetrates but the 
\ha\ photons do not. It is also possible that SNII progenitor lifetimes are simply longer than those of HII regions.\\
SNIb/c show a strong association with the \ha\ of their host galaxies, as would be expected for SN progenitors of the highest mass.
The marginal difference found when we separate these SNe into their subgroups, should be treated cautiously due to the sample size, and the probable
errors made in some of the classifications of these events.
We have recently obtained additional imaging of SNIb/c hosts that will more than double our sample size. The problem of misclassification will be addressed
in a forthcoming paper on this increased sample, along with new results and detailed analysis.\\
Due to the small number of SNIa in our sample it is hard to draw any firm conclusions on this progenitor type. It is however interesting
to speculate on the significance of our result given recent predictions \cite{mann06} that the progenitor population of SNIa 
is bimodal, with around half of the population arising from progenitors with lifetimes of $\sim$10$^{8}$yrs, 
the so called `prompt' SNe, and the other half arising from relatively longer lived
progenitors (a few Gyrs), the `tardy' SNe. One could cautiously assign those SNIa with positive NCR values 
to the `prompt' fraction. However, it is likely that some or possibly all of these `prompt' SNe in our sample
are down to chance alignments,
and more observations and analysis is needed before we draw any further conclusions.\\
We are continuing to increase our sample size of all SN types, obtaining additional \ha\ imaging with various telescopes. 
We are also now starting to obtain a wider range in wavelength coverage of our host galaxies, in particular those of SNIa, in order to map 
out a wider range in stellar ages and metallicities. We will use these observations in conjunction with the stellar population synthesis models
of Salaris and collaborators \cite{piet04}, in order to further tie down  the SNIa progenitor population.


\begin{theacknowledgments}
The JKT and the INT are operated on the island of La Palma by the Isaac Newton Group in the Spanish Observatorio 
del Roque de los Muchachos of the Instituto de Astrofisica de Canarias. 
The Liverpool Telescope is operated on the island of La Palma by Liverpool John Moores University in the Spanish 
Observatorio del Roque de los Muchachos of the Instituto de Astrofisica de Canarias with financial support from the UK Science and Technology Facilities Council.
This research has made use of the NASA/IPAC Extragalactic Database (NED), which is operated by the Jet Propulsion Laboratory,
California Institute of Technology, under contract with the National Aeronautics and Space Administration. We also thank
the Cambridge Astronomical Survey Unit for the reduction of Wide Field Camera observations with the INT.
\end{theacknowledgments}



\bibliographystyle{aipproc}   

\bibliography{reference.bib}

\IfFileExists{\jobname.bbl}{}
 {\typeout{}
  \typeout{******************************************}
  \typeout{** Please run "bibtex \jobname" to optain}
  \typeout{** the bibliography and then re-run LaTeX}
  \typeout{** twice to fix the references!}
  \typeout{******************************************}
  \typeout{}
 }

\end{document}